\documentclass[%
 pre,preprint,
 amsmath,amssymb,
 aps,
]{revtex4-1}

\usepackage{graphicx}
\usepackage[space]{grffile}
\usepackage{latexsym}
\usepackage{textcomp}
\usepackage{longtable}
\usepackage{tabulary}
\usepackage{booktabs,array,multirow}
\usepackage{amsfonts,amsmath,amssymb}
\usepackage{natbib}
\usepackage{url}
\usepackage{hyperref}
\hypersetup{colorlinks=false,pdfborder={0 0 0}}
\usepackage{etoolbox}
\makeatletter

\newif\iflatexml\latexmlfalse

\AtBeginDocument{\DeclareGraphicsExtensions{.pdf,.PDF,.eps,.EPS,.png,.PNG,.tif,.TIF,.jpg,.JPG,.jpeg,.JPEG}}

\usepackage[utf8]{inputenc}
\usepackage[german,english]{babel}
\usepackage[bottom=1in]{geometry}

\usepackage{amsfonts}

\iflatexml

\else

\begin{document}

\title{Effective potential reveals evolutionary trajectories in complex fitness
landscapes}

\author{Matteo Smerlak}
\affiliation{Max Planck Institute for Mathematics in the Sciences, Leipzig, Germany}

\selectlanguage{english}
\begin{abstract}
Growing efforts to measure fitness landscapes in molecular and microbial
systems are premised on a tight relationship between landscape
topography and evolutionary trajectories. This relationship, however, is
far from being straightforward: depending on their mutation rate,
Darwinian populations can climb the closest fitness peak (survival of
the fittest), settle in lower regions with higher mutational robustness
(survival of the flattest), or fail to adapt altogether (error
catastrophes). These bifurcations highlight that evolution does not
necessarily drive populations ``from lower peak to higher peak'', as
Wright imagined. The problem therefore remains: how exactly does a
complex landscape topography constrain evolution, and can we predict
where it will go next? Here I introduce a generalization of quasispecies
theory which identifies metastable evolutionary states as minima of an
effective potential. From this representation I derive a coarse-grained,
Markov state model of evolution, which in turn forms a basis for
evolutionary predictions across a wide range of mutation rates. Because
the effective potential is related to the ground state of a quantum
Hamiltonian, my approach could stimulate fruitful interactions between
evolutionary dynamics and quantum many-body theory.%
\end{abstract}%

\maketitle

\section*{\texorpdfstring{\emph{Significance
statement}}{Significance statement}}

{\label{874460}}

\emph{The course of evolution is determined by the relationship between
heritable types and their adaptive values, the fitness landscape. Thanks
to the explosive development of sequencing technologies, fitness
landscapes have now been measured in a diversity of systems from
molecules to micro-organisms. How can we turn these data into
evolutionary predictions? I show that preferred evolutionary
trajectories are revealed when the effects of selection and mutations
are blended in a single effective evolutionary force. With this
reformulation, the dynamics of selection and mutation becomes Markovian,
bringing a wealth of classical visualization and analysis tools to bear
on evolutionary dynamics. Among these is a coarse-graining of
evolutionary dynamics along its metastable states which greatly reduces
the complexity of the prediction problem.}

\section*{}

{\label{874460}}

\section{Introduction}

{\label{874460}}

Darwinian evolution is the motion of populations in the space of all
possible heritable types graded by their reproductive value, the fitness
landscape~\cite{Stadler,Orr_2009,Fragata_2019}. In~ Wright's vivid words, the interaction
of selection and variation enables populations to~ ``continually find
their way from lower to higher~peaks''~\cite{wright1932}, thereby
providing a universal mechanism for open-ended
evolution~\cite{de_Vladar_2017}. Thanks to the explosive development of
sequencing technologies, fitness landscapes have now been measured in a
variety of real molecular~\cite{Blanco_2019}, viral~\cite{Dolan_2018}
or microbial~\cite{de_Visser_2014} systems. As a result, the goal
of~\emph{predicting} evolution no longer appears wholly out of
reach~\cite{Weinreich_2006,Lobkovsky_2012,de_Visser_2014,L_ssig_2017,de_Visser_2018}. In essence, if we know the topography of the
fitness landscape---its peaks, valleys, ridges, etc.---we should be able
to estimate where a population is likely to move next. Making such
predictions from high-resolution fitness assays is a central challenge
of quantitative evolutionary theory.~

An array of measures quantifying the topography of fitness landscapes
has been developed in support of this program. Especially important from
the dynamical perspective, the~\emph{ruggedness} of a landscape can be
measured in a variety of ways, some simple but coarse (density of
fitness maxima~\cite{Kauffman_1987}, correlation
functions~\cite{Weinberger_1990,Stadler_1996}), others detailed but more involute
(amplitude spectra~\cite{Hordijk_1998}, geometric
shapes~\cite{Beerenwinkel_2007}). As stressed by
Kimura~\cite{Kimura_1983},~\emph{neutrality}---the distribution of
plateaus rather than peaks---is another key feature of real
landscapes~\cite{Cowperthwaite_2007} which derives from their high-dimensional
nature~\cite{MAYNARD_SMITH_1970,Gavrilets_1997} and can be studied with the tools of
percolation theory~\cite{Reidys_1997}. The NK(p)~\cite{Kauffman_1989,barnett1998},
Rough Mount Fuji~\cite{Aita_2000,Neidhart_2014}, holey~~\cite{MAYNARD_SMITH_1970,Gavrilets_1997} and
other models in turn provide simple landscapes with tunable ruggedness
and/or neutrality, which can be used to fit empirical data and simulate
evolutionary trajectories. On these foundations a new subfield of
mathematical biology has emerged, the quantitative analysis of fitness
landscapes~\cite{Szendro_2013}.~

What these fitness-centric measures fail to capture, however, is the
fact that~\emph{populations with different mutation rates experience the
same fitness landscape differently}. This is already clear if we
consider the rate of fitness valley crossings, which strongly depends on
the mutation rate~\cite{VANNIMWEGEN_2000,Weissman_2009} and therefore cannot be computed
from topographic data alone. But Eigen's quasispecies theory showed that
varying mutation rates can also have a~\emph{qualitative} effect on
evolutionary trajectories, potentially leading to error catastrophes and
the loss of adaptation~\cite{Eigen_1971}. More subtly, mutational
robustness has been shown to evolve neutrally~\cite{van_Nimwegen_1999} and to
sometimes outweigh reproductive rate as a determinant of evolutionary
success (``survival of the flattest'')~~\cite{Wilke_2001,Codo_er_2006}. These
evolutionary bifurcations are not mere theoretical curiosities: lethal
mutagenesis---an effort to push a population beyond its error
threshold---is a promising therapeutic strategy against certain viral
pathogens~\cite{Eigen_2002,Domingo_2019} and perhaps cancer \cite{Sol__2004}.~

More fundamentally, these bifurcations imply that, unless mutations are
so rare that populations are genetically homogeneous at all times and
selection is~ periodic (the so-called strong-selection weak-mutation
(SSWM) limit~\cite{Gillespie_1983}), evolving populations are not
necessarily attracted to fitness peaks. The SSWM assumptions are
typically violated in large microbial populations, where multiple
mutants often compete for fixation in a process known as clonal
interference~~\cite{Gerrish_1998,Park_2007}. In yet stronger mutation regimes,
e.g. in RNA viruses, natural selection acts on clouds of genetically
related mutants rather than on individuals, and evolution is the
intermittent succession of stabilization-destabilization transitions for
these clouds~\cite{Eigen_2007}. In the presence of neutrality, epochal
or ``punctuated'' evolution can also arise through the succession of
slow intra-network neutral diffusion and fast inter-network
sweeps~\cite{Huynen_1996}.

These results raise fundamental questions regarding the~\emph{dynamical}
analysis of fitness landscapes: When is flatter better than fitter?
Where are the evolutionary attractors in a given landscape with
ruggedness and/or neutrality? What quantity do evolving populations
optimize? Can we estimate the time scale before another attractor is
visited? More simply, can we predict the future trajectory of an
evolving population from its current location, the topography of its
landscape, and the mutation rate?~

In this paper I outline a mathematical framework to address these
questions in large, asexual populations, for both genotypic (discrete,
high-dimensional) and phenotypic (continuous, low-dimensional) fitness
landscapes. Inspired by certain analogies with the physics of disordered
systems, I show that the selection-mutation process can be understood as
a random walk or diffusion in an effective potential---the same kind of
dynamics as, say, protein folding kinetics~\cite{Bryngelson_1995}. This
representation reduces the~\emph{a priori}~difficult problem of
identifying evolutionary attractors and dominant trajectories in a
complex fitness landscape to the much more familiar problem of Markovian
metastability~\cite{H_nggi_1990}.~In contrast with another classical
Markovian model of evolution, Gillespie's adaptive walk
model~\cite{Gillespie_1983,Kauffman_1987}, my approach is not restricted to the SSWM
regime and fully accommodates genotypic and/or phenotypic heterogeneity
in evolving populations~\cite{Gerrish_1998,Park_2007}. Moreover, because the
effective potential integrates fitness and mutational robustness in a
single function on the space of types, it is also more suited to
analyze---and eventually predict---the dynamics of a population than the
bare fitness landscape from which it derives.~

\section{Results}

{\label{201174}}

\subsection{Selection-mutation
dynamics}

{\label{847231}}

Consider a fitness landscape $\Phi=(X,\Delta,\phi)$, consisting of a space of types $X$, a mutation operator $\Delta$ on $X$ and a (Malthusian) fitness function $\phi:X\rightarrow \mathbb{R}$. The nature of the landscape is left unspecified: $\Phi$ could be a be genotypic landscape, in which case $X$ will be a finite graph (usually a hypercube or some more general Hamming graph), and  $\Delta$ its Laplacian matrix; or $\Phi$ could be a quantitative phenotypic landscapes, and then $X$ will be a domain of $\mathbb{R}^d$ and $\Delta$ a differential operator thereon, usually the Laplacian (if mutational effects are sufficiently small and frequent). 

We assume a large asexual population evolving on this landscape according to the continuous-time Crow-Kimura \cite{J__1971} selection-mutation equation 
\begin{equation}\label{crow_kimura}
\frac{\partial p_t(x)}{\partial t}=\big(\phi(x)-\langle \phi\rangle_t\big)p_t(x)+\mu \Delta p_t(x),
\end{equation}
where $p_t(x)$ is the distribution of types $x\in X$ at time $t$, $\langle \phi\rangle_t$ the population mean fitness, and $\mu$ the mutation rate per individual per unit time. In contrast with previous analytical works which focused on finding exact solutions to \eqref{crow_kimura} \cite{BAAKE_2001}, our goal is to understand the motion of the distribution $p_t(x)$ in the landscape without making restrictive assumptions on its topography. This is necessary for the predictive analysis of real fitness landscapes, which do not have the symmetries of soluble models. 

Note that \eqref{crow_kimura} assumes that mutations occur independently of replication events. The results in this paper do not depend on this assumption: we could equally well consider Eigen's quasispecies model \cite{Eigen_2007}, where mutations only arise as replication errors, or more general models (Methods). On the other hand, the infinite population limit implicit in \eqref{crow_kimura} is a real limitation which overlooks the stochasticity of evolutionary histories. The applicability of deterministic models has been discussed extensively in the literature \cite{Eigen_2007,Wilke_2005}, including from an experimental perspective \cite{Domingo_2015}. Generally speaking, the infinite population approximation is good if the type space is low-dimensional or if the population is known to be localized in a small region of an otherwise high-dimensional genotypic landscape \cite{Jain_2006}. My own view is that deterministic models are a stepping stone towards any quantitative theory of evolutionary dynamics---we must understand how selection and mutation interact before we can ask about the influence of other evolutionary forces such as genetic drift. 

\subsection{Classical analytical
approaches}

{\label{663716}}

The first step to study \eqref{crow_kimura} is to linearize it. This is commonly done by simply dropping the mean fitness term $\langle \phi\rangle_t$, the probability distribution $p_t(x)$ being then obtained from the solution $f_t$ of 
\begin{equation}\label{PAM}
\frac{\partial f_t(x)}{\partial t}=\big(\mu \Delta + \phi(x)\big)f_t(x),
\end{equation}
through normalization, \textit{i.e.} $p_t=\overline{f_t}$ with $\overline{f}=f/\sum_x f(x)$. The linear equation \eqref{PAM} can then be solved formally in one of two classical ways---neither of which turns out to be directly useful for the prediction problem. 

The first common approach to \eqref{PAM} uses the Feynman-Kac formula to write $f_t(x)$ as a weighted sum over Brownian paths $X_t$ \cite{Zel_dovich_1987}. Unfortunately, these paths cover the whole fitness landscape, \textit{i.e.} they are not by themselves predictive. Alternatively, we can decompose $f_t(x)$ over a basis of normal modes of the operator $A=\mu\Delta + \phi$ and consider the evolution of each component independently, as proposed by Eigen and Schuster \cite{Eigen_2007}. This reduces \eqref{PAM} to a set of uncoupled growth equations, with the eigenvalues of $A$ as growth rates. Accordingly, evolution is seemingly reduced to the natural selection of clouds of genetically related mutants, or ``clans". 

The problem with the spectral approach is that, of all the modes of $A$, only one is positive and can therefore be interpreted as a distribution---the ``quasispecies" distribution $Q$, i.e. the eigenfunction of $A$ with the largest eigenvalue $\Lambda$. As a consequence, the Eigen-Schuster spectral approach is useful to characterize the asymptotic selection-mutation equilibrium $Q= \lim_{t\to\infty}p_t$, and in particular to determine whether this equilibrium is localized (adaptive) or delocalized (error catastrophe), but it cannot help us understand the approach to that equilibrium. 

\subsection{Effective potential
landscape}

{\label{663716}}

The key observation of this paper is that knowing $Q$---a single eigenfunction of $A$---to a good accuracy in fact goes a long way toward understanding evolutionary dynamics \textit{far} from selection-mutation equilibrium. This is because from $Q$ we can perform a change of variable that dramatically simplifies the analysis of evolutionary dynamics, as follows. Consider the function $g_t(x)=e^{-\Lambda t}Q(x)f_t(x)$, from which we can reconstruct the original distribution via $p_t(x)=\overline{Q(x)^{-1}g_t(x)}$. This function evolves according to 
\begin{equation}\label{markov}
\frac{\partial g_t(x)}{\partial t} = Lg_t(x) \quad\textrm{with}\quad L = \textrm{diag}(Q)\,(A-\Lambda)\,\textrm{diag}(Q)^{-1}.
\end{equation}
In Methods I show that, for any operator $A$ that preserves the positivity of distributions, \eqref{markov} is the forward Kolmogorov equation of a reversible \textit{Markov process} with effective potential
\begin{equation}
U(x)=-2\log Q(x).
\end{equation}
In the case where $\Delta$ is the Laplacian operator this process is just a biased random walk/Brownian motion:
\begin{itemize}
\item 
For discrete types, $L$ generates nearest-neighbor jumps with transition rate 
$$L_{x\to y}=\mu \, \exp\left(-\frac{U(y)-U(x)}{2}\right).$$ 
\item 
For continuous types, $L$ is the Fokker-Planck operator for a diffusion in the potential $U$, i.e. $$Lq=-\nabla\cdot j \quad\textrm{with}\quad j= \mu\big(-\nabla q - q\nabla U\big).$$
\end{itemize}

Note that the interpretation of the derived Markov process departs from that of the original selection-mutation model in two ways. First, $Q$ is no longer viewed as coding the asymptotic equilibrium between selection and mutation, in which all transients are washed out; instead, (two times minus) its logarithm acts a potential landscape, whose role is to prescribe the dynamics \textit{away} from equilibrium. Second, we are used to thinking of mutations as adding a random component to the otherwise deterministic flow of natural selection, with $\mu$ controlling the strength of the noise. Here, by contrast, $\mu$ plays the role of $(i)$ an (inverse) time scale, and $(ii)$ a parameter of the effective potential $U$ which directs the evolution of the density in the space of types $X$. The noise component of the process itself has \textit{unit} diffusivity. 

What is the benefit of replacing the selection-mutation operator $A$ by the Markov generator $L$? The answer is that the latter  has an inbuilt notion of dominant evolutionary trajectory: from a given type $x$, the preferred path is the line of steepest descent of the effective potential $U$. Moreover, thanks to the smoothing effect of mutations imprinted in the quasispecies distribution, the potential landscape is far simpler---in particular, less rugged---than the fitness landscape itself. We now illustrate these aspects in more detail.  

\subsection{Bare vs. effective
ruggedness}\label{bare-vs.-effective-ruggedness}

As already mentioned, a classic approach to the ruggedness of fitness landscapes consists in counting the number of local fitness maxima \cite{Kauffman_1987}. For instance, in $NK$ landscapes the expected density of fitness peaks grows from $2^{-N}$ (additive or ``Mount Fuji" landscape) to $(N+1)^{-1}$ (uncorrelated or ``house of cards" landscape) as the epistasis parameter $K$ increases from $0$ to $N-1$, irrespective of the distribution of fitness components. However, the number of fitness peaks---the \textit{bare} ruggedness of the landscape---is not directly relevant for evolutionary trajectories: at finite mutation rates, a low peak can be indistinguishable from no peak.

The results in the previous section imply that, in the large population limit, the evolutionary attractors are the local maxima of $Q$ (local minima of $U$), not those of $\phi$. But for a type $x$ to be a local maximum of $Q$, it is not enough that its fitness be greater than that of its one-step mutants: an approximate expression for $Q$ (given in \eqref{FWA} below and valid at low $\mu$) shows that $\phi(x)$ must instead be greater than $\max \phi-\mu$. This condition is typically much more stringent than the condition for $x$ to be a local fitness maximum; the effective potential landscape is therefore significantly smoother than the fitness landscape. Thus, the number of $Q$-maxima of an NK lansdcape does not actually increase with $K$, but does with the skewness of the distribution of fitness components (data not shown).  

\subsection{Reduced evolutionary
dynamics}\label{reduced-evolutionary-dynamics}

Next, the Markovian reformulation immediately suggests a coarse-grained (reduced) representation of evolutionary dynamics, as follows. For each local minimum $x_\alpha$ of $U$ we can consider the set of types $X_\alpha$ from which $x_\alpha$ can be reached along a $U$-decreasing path, its basin of attraction. The potential barrier between two adjacent basins is then given by $B_{\alpha\to \beta}=\min_{\pi} \max_{x\in \pi} [U(x)-U(x_\alpha)]$ where $\pi$ spans the directed paths connecting $X_\alpha$ to $X_\beta$. According to the standard Arrhenius-Kramers law for the transition time between minima of a potential landscape \cite{H_nggi_1990}, the basin $X_\alpha$ with frequency $\sum_{x\in X_\alpha}p_t(x)$ is \textit{metastable} if 
\begin{equation}
\min_\beta B_{\alpha\to\beta}\gg 1.
\end{equation}
Large deviation theory further indicates that, of all the possible escapes from $X_\alpha$ to an adjacent basin, the transition to $\textrm{argmin}_\beta B_{\alpha\to\beta}$ is exponentially more likely to happen. This reduction in dynamical complexity is the main result of this paper.  

The coarse-grained dynamics can be represented using tools usually applied to energy landscapes, such as the basin hopping graphs (BHG) recently developed in the context of RNA folding \cite{Kuchar_k_2014}. In a nutshell, a BHG is obtained by collapsing the local minima $x_\alpha$ and their basins of attraction $X_\alpha$ into nodes and connecting them according to adjacency relations between basins, weighted by the barrier height $B_{\alpha\to\beta}$, as in Fig. \ref{793071}. 

\subsection{An evolutionary Lyapunov
function}

{\label{850652}}

Finally, the Markovian reformulation provides a novel Lyapunov function for selection-mutation dynamics. An evolutionary Lyapunov function (ELF) traditionally refers to one of two distinct concepts. The first notion of ELF is a monotonic functional of distributions over type space $X$; examples include Fisher's variance functional in the pure selection regime \cite{Fisher_1930} or for type-independent mutation rates \cite{sigmund1998}, or Sella and Hirsh's free fitness functional in the SSWM regime \cite{Sella_2005} (see also \cite{Jones_1978}). The second kind of ELF is a monotonic functional of distributions over distributions over type space $X$ (\textit{i.e.} over allele frequency distributions); Iwasa's \cite{Iwasa_1988} and Mustonen and L\"assig's \cite{Mustonen_2010} free fitness functions are of this kind.

Here I introduced a Markovian version of evolutionary dynamics in type space which is not restricted to pure selection or SSWM regimes. Since this Markov processes is reversible, the relative entropy (or Kullback-Leibler divergence) $D[\,\cdot\, \Vert\, \cdot\,]$ with respect to its equilibrium distribution $\propto e^{-U}=Q^2$ must decreases monotonically in time. This means that
\begin{equation}\label{lyapunov}
F[p_t]=D[\overline{Qp_t}\,\Vert \,\overline{Q^2}]
\end{equation}
is a Lyapunov function for the evolutionary equation \eqref{crow_kimura} for any mutation operator $\Delta$ and any mutation rate $\mu$ (Fig. \ref{970511}). The construction of this ELF follows the same pattern as Iwasa's and Mustonen and L\"assig's (as a relative entropy), but, unlike theirs but like Fisher's, results in a functional of distributions over $X$ and not allele frequency space. Also note that $F[p_t]$ is not merely an additive correction to mean fitness and thus goes beyond the scope of ``free fitness" functions. 

\section{Examples}

{\label{755862}}

To illustrate the predictive value of the Markovian formulation of
selection-mutation dynamics we now consider two simulated fitness
landscapes, chosen such that evolutionary attractors are not easily read
off the landscape itself. (See Methods for explicit definitions.)

\subsection{Two-dimensional lattice}

{\label{816649}}

\begin{figure}[h!]
\begin{center}
\includegraphics[width=\textwidth]{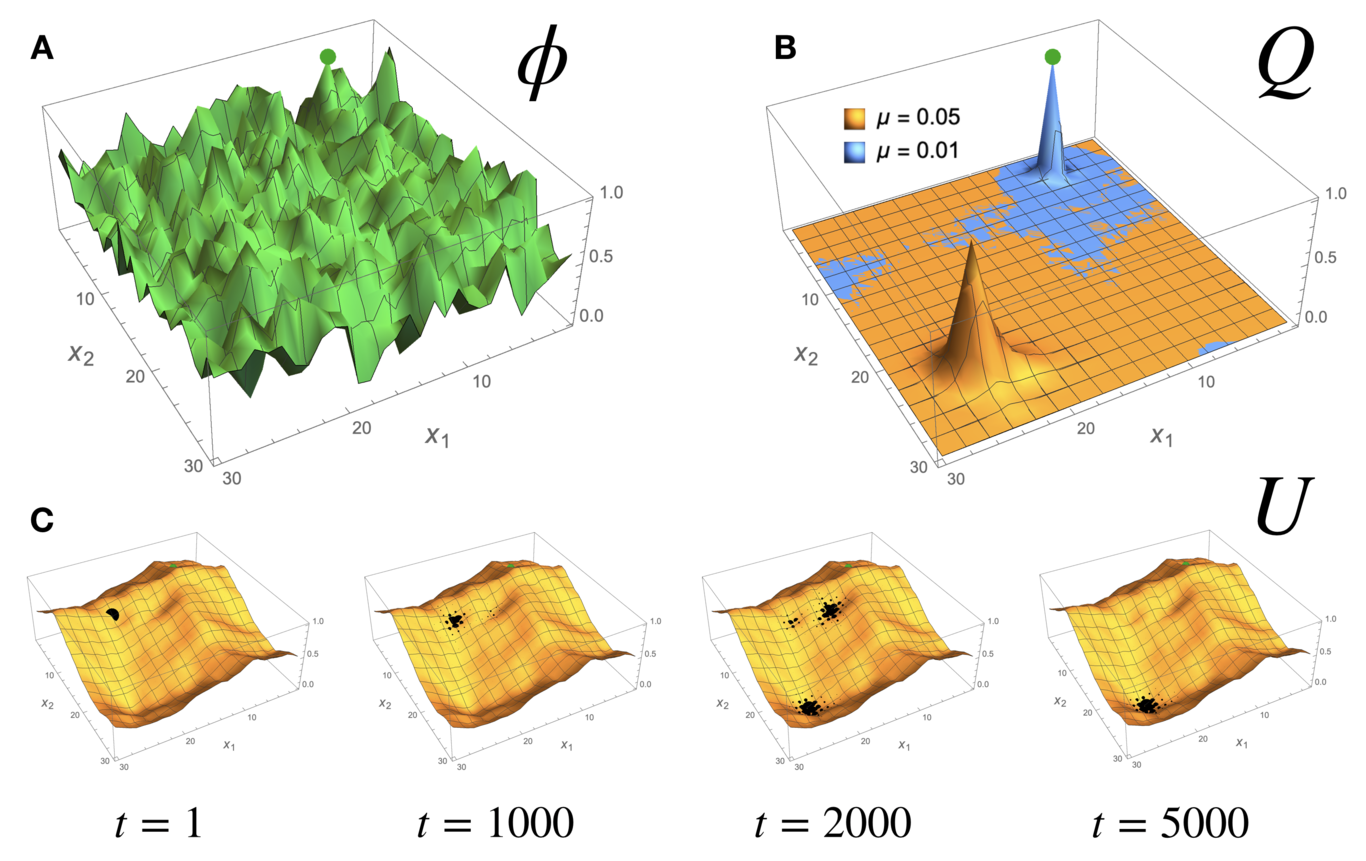}
\caption{{Evolution in a rugged 2d fitness landscape. A: The fitness landscape,
obtained by sampling a Gaussian process with unit standard deviation and
unit correlation length; the global fitness maximum is indicated by the
green dot. It is~\emph{a priori} difficult to predict the path taken by
a population evolving in this landscape. B: The quasispecies
distributions~\(Q\) for two different values of the
mutation rate~\(\mu\), localized at the fitness peak
(low~\(\mu\)) or in some lower but flatter region
(high~\(\mu\)). C: The effective potential~\(U=-2\log Q\)
for~\(\mu\ =\ 0.05\) is much smoother than the fitness landscape, with
few local minima which act as local attractors for an evolving
population (black dots). Note how the population conspicuously moves
away from the global fitness maximum.
{\label{842686}}%
}}
\end{center}
\end{figure}

 We begin with a two-dimensional rugged ``phenotypic"\footnote{Phenotypic landscapes are not expected to be rugged the way genotypic landscapes are, at least not in a biological context. I chose this example for the ease of its visualization, as well as for its connection with Anderson localization.} landscape, generated by sampling values from a Gaussian process with unit correlation length on a $30\times 30$ lattice (with periodic boundary conditions). In the realization shown in Fig. \ref{842686}A, the fitness landscape has a unique global maximum (green dot); this type corresponds to the maximum of the quasispecies $Q$ for $\mu \leq 0.02$ (survival of the fittest), but not for higher mutation rates (survival of the flattest), see Fig. \ref{842686}B. 

Predicting the evolution of an initially monomorphic population directly from the topography of $\phi$ is clearly a difficult proposition. By contrast, examination of the effective potential $U=-2\log Q$ (Fig. \ref{842686}C) immediately reveals the preferred directions for its evolution: the population will go downhill in the potential $U$, potentially getting transiently trapped in the basins of its local minima and making transitions to other basins along the lowest saddles separating them. This is indeed the behavior of numerical solutions of the Crow-Kimura equation (Fig. \ref{842686}C).\selectlanguage{english}

\subsection{Binary sequences with
neutrality}

{\label{318692}}

\begin{figure}[h!]
\begin{center}
\includegraphics[width=\textwidth]{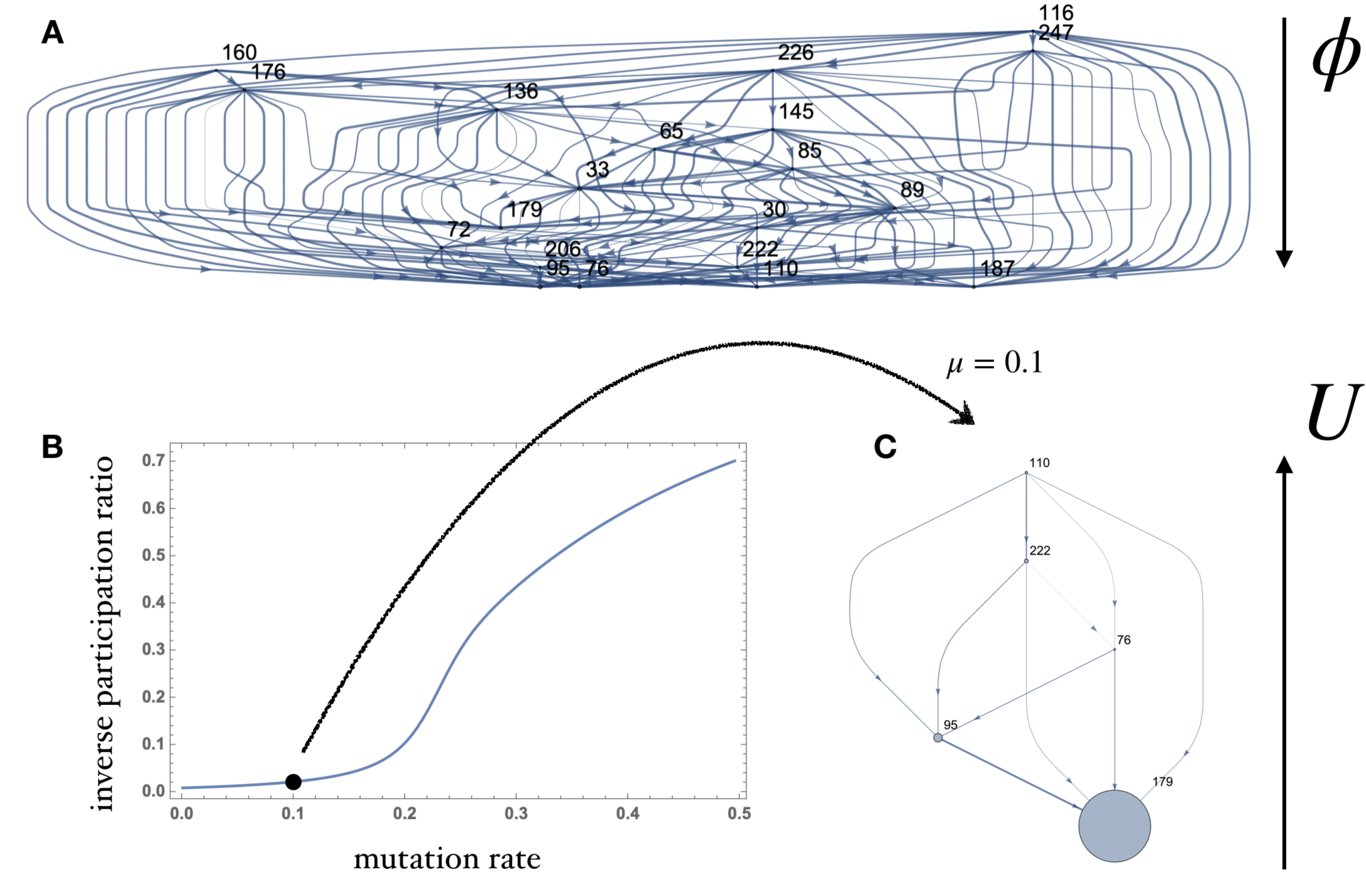}
\caption{{Evolution in an~\emph{NKp} genotypic landscape with~\(2^8=256\)
types. A: The fitness landscape has~\(20\)~local fitness
maxima and many saddles between them, making visualization and
evolutionary prediction challenging. Here the landscape is represented
as a basin hopping graph (BHG), in which nodes are basins of attractions
of fitness maxima and edges adjacency relations between basins weighted
by the barrier height. B: As the mutation rate passes a threshold
at~\(\mu\simeq0.2\) (in units of the maximal fitness difference), the
quasispecies distribution delocalizes, as signalled by the inverse
participation ratio~\(\left(\sum_x Q(x)^2\right)^{-1}/\vert X\vert\). C: The BHG for the effective
potential~(here for~\(\mu = 0.1\)) is much simpler---and
immediately predictive, see Fig.~{\ref{970511}}.~
{\label{793071}}%
}}
\end{center}
\end{figure}

\begin{figure}[h!]
\begin{center}
\includegraphics[width=\textwidth]{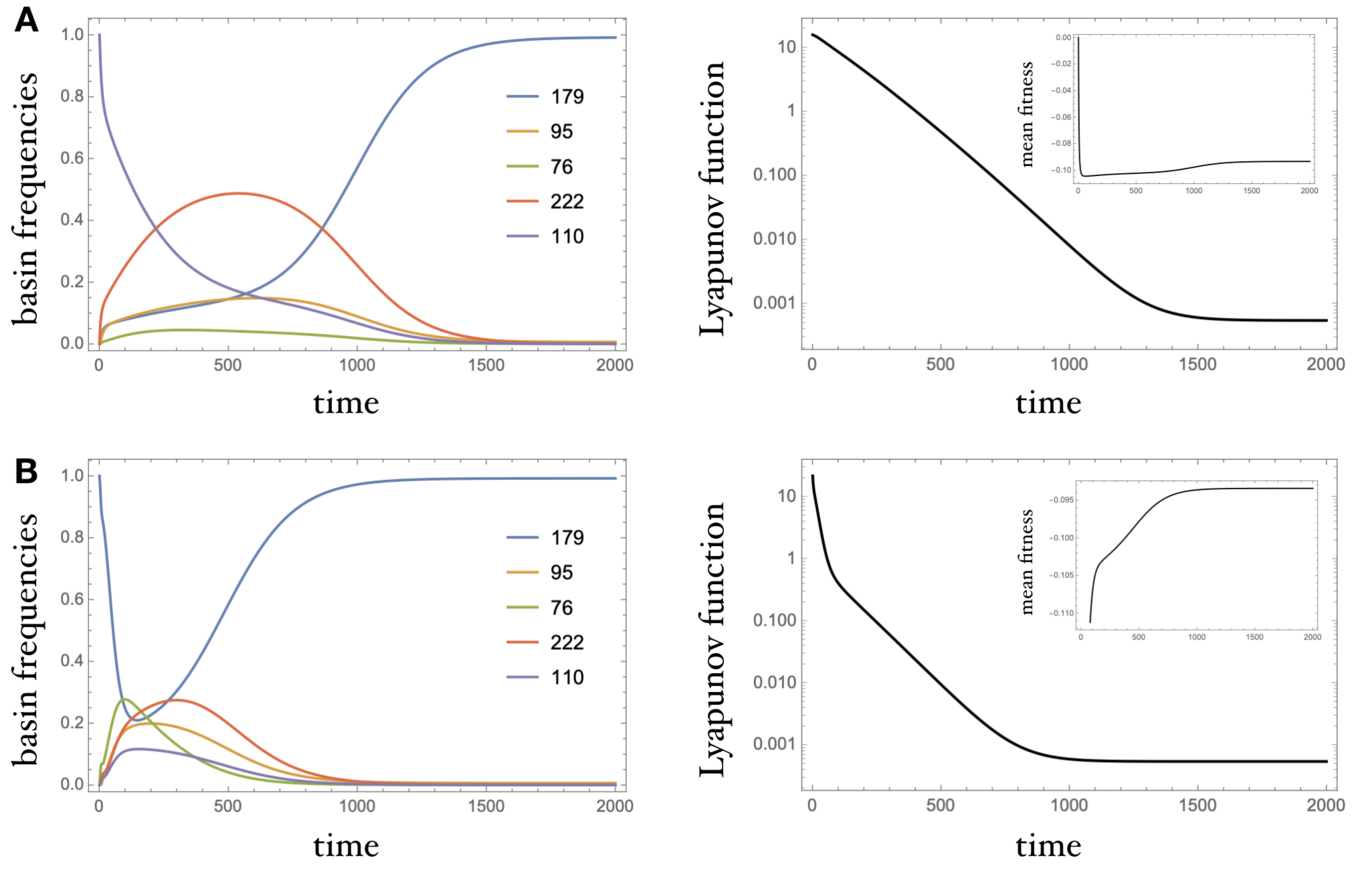}
\caption{{Evolutionary trajectories in the landscape of
Fig.~{\ref{793071}}A, obtained by integration of the
Crow-Kimura equation. A: A population initially concentrated in basin
110 moves towards basin 179 through basins 222 and 95, as suggested by
the BHG in Fig.~{\ref{793071}}C. This happens in spite
of the fact that 110 is a global fitness maximum and mean fitness
decreases in time.~ B: Here the population starts off concentrated at
type 179 and spreads in other basins under the effect of mutations,
before returning to the basin of 179 as \(t\to\infty\). This
non-monotonic behavior of the basin frequency does not prevent the
evolutionary Lyapunov function to decrease monotonically.~
{\label{970511}}%
}}
\end{center}
\end{figure}

As a simple model of a genotypic landscape with both ruggedness and neutrality, I considered an $NKp$ landscape \cite{barnett1998a} of binary sequences with length $N=8$, epistasis parameter $K=6$ and neutrality parameter $p=0.7$ (details in Methods). The landscape in Fig. \ref{793071} has $20$ local maxima and an error threshold at $\mu_c\simeq 0.2$. Comparing the basin hopping graphs of the fitness landscape $\phi$ and of the potential landscape $U$ reveals that most of the complexity of the former is spurious. Moreover, coarse-grained evolutionary trajectories, described by the basin frequencies $p(X_\alpha)$, is consistent with the succession of transitions predicated by the basin hopping graph of $U$: a population initially concentrated around the genotype $110$ (a global fitness maximum) will evolve towards the flatter genotype $179$ via the basins of $222$ and $95$  (Fig. \ref{970511}A). 

One also checks that the Lyapunov function \eqref{lyapunov} decreases monotonically also when the mean fitness $\langle\phi\rangle_t$ does not (Fig. \ref{970511}A) and when the basin frequencies have strongly non-monotonic behavior (Fig. \ref{970511}B).\selectlanguage{german} \selectlanguage{english}
\selectlanguage{english}

\section{Physical analogies}

{\label{459866}}

Evolutionary theory has long benefited from analogies with statistical
physics---the other field of science dealing with large, evolving
populations---, see~\emph{e.g.}~\cite{Sella_2005,Mustonen_2010,de_Vladar_2011,Smerlak_2017}. More recently,
Leuth\selectlanguage{german}äusser~\cite{Leuth_usser_1986} and others\emph{~}\cite{Baake_1997,Saakian_2004}~have
highlighted a parallel between evolutionary models in genotype space and
certain~\emph{quantum} spin systems, which can be leveraged to compute
the quasispecies distribution~\(Q\) for some special
fitness landscapes~\cite{BAAKE_2001}. The present work was inspired by
the realization that, in quantum mechanics, it~\emph{is} possible to map
Schrödinger operators to diffusive trajectories---this is the basic idea
underlying Nelson's ``stochastic'' formulation of quantum~
mechanics~\cite{Nelson_1966}.~

But the scope of the analogy between evolution and non-equilibrium
physics is, in fact, much broader: the interplay between selection and
mutation is typical of~\emph{localization phenomena in disordered
systems~\cite{Stollmann_2001}}, be them classical or quantum. The
linearized Crow-Kimura equation {\ref{PAM}}, for
instance, is formally identical to the parabolic Anderson model
~\cite{Zel_dovich_1987,Carmona_1994,K_nig_2016}, a simple model of~intermittency in random fluid
flows; the linearized Eigen model in turn resembles the Bouchaud trap
model~\cite{Bouchaud_1992}, a classical model of slow dynamics and ageing
in glassy systems. These physical phenomena have obvious evolutionary
counterparts: the Anderson localization transition corresponds to the
error threshold; intermittency to epochal or punctuated evolution;
tunnelling instantons to fitness valley crossings; and ageing to
diminishing-return epistasis.~The generalization of Nelson's mapping of
the Scrödinger equation to a diffusion process presented in this paper
implies that all are in fact unified under the familiar umbrella of
Markovian metastability.

\begin{figure}[h!]
\begin{center}
\includegraphics[width=\textwidth]{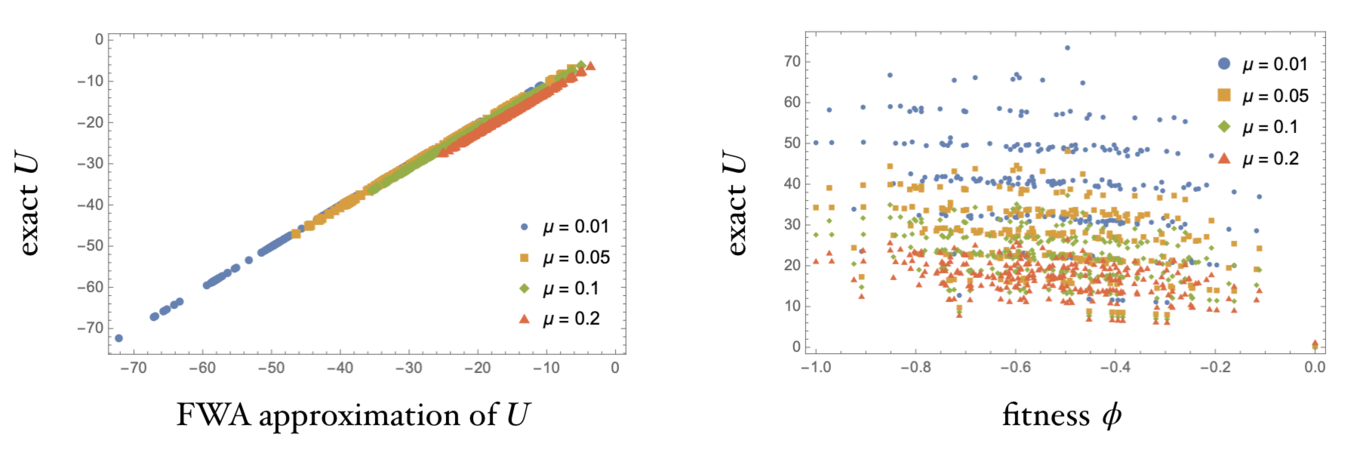}
\caption{{Effective potential for a non-degenerate NK landscape
with~\(N=8\) and~\(K=6\). The FWA approximation
familiar from Anderson localization theory gives excellent results,
including at large mutation rates (left). By contrast, the bare fitness
values~\(\phi\) are poorly correlated with the effective
potential~\(U\) (right). Here mutation rates are given in
units such that~\(\phi\) ranges from\(-1\)
to~\(0\). -
{\label{982486}}%
}}
\end{center}
\end{figure}

The value of such analogies is twofold. First, they bring the large
repertoire of results and techniques derived in condensed matter and
nonequilibrium physics to bear on evolutionary dynamics. As an
elementary example, we can use the forward scattering approximation
(FWA) familiar from Anderson and many-body localization
theory~\cite{Pietracaprina_2016}~to compute the effective
potential~\(U\) in the small~\(\mu\)~limit, as
the (log of the) ground state of the quantum
Hamiltonian~\(H\ =-A\ =\ -\mu\Delta-\ \phi\); for non-degenerate fitness landscapes
this gives~

\begin{equation}\label{FWA}
U(x)\underset{\mu\to 0}{\sim}U(x_*)-2\log \sum_{\pi}\prod_{i\in\pi} \frac{\mu}{\phi(x_*)-\phi(\pi_i)},
\end{equation}
where $x_*=\textrm{argmax}\,\phi$ and $\pi$ spans the set of shortest paths between $x_*$ and $x$. This gives suprisingly good results, including for large mutation rates (Fig. \ref{982486}).\footnote{Note that \eqref{FWA} may be interpreted as showing that, whether or not the fitness landscape is correlated, the effective potential has the topography of a Rough Mount Fuji with slope $2\log(\mu/\sigma_\phi)$ where $\sigma_\phi$ is the range of fitness values.} Conversely, the link between evolution and the physics of disordered media can stimulate new work in physics and mathematics. As already mentioned, the generator of selection-diffusion dynamics is not always Hermitian (it is not in Eigen's model). This suggests that some of the  results usually derived for random Schrödinger operators can likely be generalized for more general classes classes of operators, as already emphasized by Altenberg \cite{Altenberg_2012}.\selectlanguage{english}

\section{Conclusion}

{\label{459866}}

A widely shared understanding of the role of mutations in evolution has
them feeding raw material to the fitness-maximizing sieve of natural
selection. But when mutation rates are high, as they are
in~\emph{e.g.~}RNA viruses~\cite{Drake_1999} and likely were in early
life~\cite{Eigen_2007}, evolutionary success requires more than the
discovery of a high-fitness mutant genotype: the mutants of the new
mutant must also have relatively high fitness,~\emph{i.e.} the mutant
type must be mutationally robust. The effective
potential~\(U\) introduced in this paper combines fitness
and flatness into a single evolutionary potential---should we call it~
``flitness''?---which directly determines evolutionary trajectories
across the spectrum of mutation rates. I argue that instead of the
fitness landscape itself, it is this effective potential that we should
analyze, coarse-grain, etc. if we are to predict evolution.~~

On a conceptual level, the effective potential~\(U\)
addresses two longstanding questions in evolution:~\emph{(i)~}On what
time scale (individual generation, infinite lineage) should ``fitness''
be defined~\cite{bouchard2015}? and~\emph{(ii)} What quantity does
evolution optimize~\cite{Smith_1978}? My proposed answers are,
respectively:~\emph{(i)} It is fine to define the
fitness\(\ \phi\left(g\right)\) of a type~\(g\) as reproductive
success over one generation, which makes it directly measurable, but one
should keep in mind that\(\ \phi\left(g\right)\) is not necessarily a good
predictor for the success of a lineage descending
from~\(g\)---this role is played by the effective
potential~\(U\left(g\right)\); and~\emph{(ii)} like other dissipative
processes, evolution through selection and mutations minimizes the
statistical divergence to its Markovian equilibrium. There is an arrow
of time in micro-evolution---just not one that points towards maximal
fitness.

\par\null

\section*{Acknowledgements}

{\label{728266}}

This work was inspired by the results of Filoche and Mayboroda on wave
localization~\cite{Filoche_2012} and those of Yasue on quantum
tunnelling~\cite{Yasue_1978}. I am indebted to M. Kenmoe, A. Klimek, M.
L\selectlanguage{german}ässig, O. Rivoire and D. Saakian for discussions and feedback. Funding
for this work was provided by the Alexander von Humboldt Foundation in
the framework of the Sofja Kovalevskaja Award endowed by the German
Federal Ministry of Education and Research.

\section*{Methods}

{\label{750397}}

\subsection*{From positive to Markov
semigroups}

{\label{540167}}

The main result of this paper is best formulated in terms of positive operator semigroups \cite{rhandi2017}. A positive operator semigroup $(P_t)_{t\geq 0}$ is one that preserves the positivity of distributions on a space $X$, but not their normalization. This is the case of the linear flow $(P_t)=(e^{At})$ if the non-diagonal elements of its generator $A$ are all non-negative (i.e. $A$ is ``essentially non-negative"). Up to the addition of a multiple of the identity, we may further assume that the diagonal elements are also non-negative, i.e. $A$ is a non-negative operator.

The Perron-Frobenius theorem states that $A$ has a left eigenvector $Q$ with simple eigenvalue $\Lambda$ whose components are all positive in each irreducible component; moreover $P_t=e^{At}$ converges to the projection operator on $Q$ as $t\to\infty$. Now, under the conditions above, the operator 
\begin{equation}
L=\textrm{diag}(Q)\,(A-\Lambda)\,\textrm{diag}(Q)^{-1}
\end{equation}
is the infinitesimal generator of a reversible Markov process on $X$ with equilibrium distribution $\propto e^{-U}$ with $U=-2\log Q$. This is easily proved as follows.  

If $X$ is a discrete space (genotypic landscape), we must check that $L$ satisfies the conditions for a transition rates matrix, namely that $L$ has non-negative off-diagonal elements and $\sum_{i}L_{ij}=0$. The former follows from the same property for $A$ because $L_{ij}=Q_iA_{ij}Q^{-1}_j$ for $i\neq j$. The latter follows from $Q$ being a left eigenvector of $A$ with eigenvalue $\Lambda$:
\begin{equation}
\sum_i L_{ij}=\sum_i Q_iA_{ij} Q_j^{-1}-\Lambda= \Lambda - \Lambda=0.
\end{equation}
Note that, when $A=\mu \Delta + \phi$ with $\Delta$ the Laplacian on a graph (such that $\Delta_{ij}=1$ when $i$ and $j$ are adjacent and zero if $d(i,j)>1$), then $L$ generates nearest-neighbor jumps with rate $L_{j\to i}=L_{ij}=\mu Q_iQ_j^{-1}=\mu \exp[(U_i-U_j)/2)]$, as stated in the main text.  

For the continuous case, consider a domain of $\mathbb{R}^d$ and assume for simplicity that the mutation operator $\Delta=\nabla^2$ is the Laplacian in that domain, generating a standard $d$-dimensional Brownian motion. In this way $A$ is a self-adjoint Schr\"odinger operator. Let $g_t=Qf_t^{\Lambda}$, where $\partial_t f_t^{\Lambda}=(A-\Lambda)f_t^{\Lambda}$. An explicit computation then shows that $g_t$ satisfies the continuity equation $\partial_t g_t=-\nabla\cdot j_t$ with the reversible flux $j_t=-\mu( \nabla g_t+g_t\nabla U)$. This is the Fokker-Planck equation for a diffusion process with unit diffusivity and potential $U$.

\subsection*{Model landscapes}

{\label{119668}}

The Gaussian process landscape of Fig.~{\ref{842686}}
is obtained by sampling a vector from the multivariate Gaussian
distribution with zero mean and~\(L^2\times L^2\) covariance
matrix~\(G_{x,y}=e^{-d(x,y)}\) where~\(d\) denotes the distance
function on the two-dimensional periodic lattice~~\(\mathbb{Z}_L\times\mathbb{Z}_L\).~

The $NKp$ fitness landscape over the hypercube $\{0,1\}^N$ with epistasis (or ruggedness) parameter $K$, neutrality parameter $p$ and component distribution \(\mathcal{D}\) is defined by the formula \(\phi\left(x\right)=-\frac{1}{N}\sum_{_{i=1}}^Nf_i\left(x_i,\ x_{i+1},\cdots,\ x_{i+K}\right)b_i\left(x_i,\ x_{i+1},\cdots,\ x_{i+K}\right)\) where the components of the binary string \(x\) are identified cyclically and the values of functions \(f_i,b_i:\{0,1\}^{K+1}\to\mathbb{R}\) are i.i.d. samples from \(\mathcal{D}\) and \(\textrm{Bernoulli}(1-p)\), respectively. Unless specified otherwise it is customary to take \(\mathcal{D}=\textrm{Uniform}(0,1)\). The $NK$ model is the special case when $p=0$, \textit{i.e.} without neutrality.

\selectlanguage{english}
\bibliographystyle{apsrev4-1}
\bibliography{converted_to_latex.bib%
}

\end{document}